\pdfoutput=1
\documentclass[conference]{IEEEtran}
\IEEEoverridecommandlockouts
\usepackage{cite}
\usepackage{amsmath,amssymb,amsfonts}
\usepackage{algorithmic}
\usepackage{graphicx}
\usepackage{url}
\usepackage{siunitx}
\usepackage{multicol}
\usepackage{multirow}
\usepackage{booktabs}
\usepackage{float}
\usepackage{multirow}
\usepackage{tabularx}
\usepackage{siunitx}
\usepackage{textcomp}
\usepackage{xcolor}
\usepackage[colorlinks=true, citecolor=blue, bookmarks=false]{hyperref}

\usepackage{subfigure}
\usepackage{caption}
\usepackage{cleveref}
\usepackage{algorithm}
\usepackage{algorithmic}

\usepackage[
 activate={true,nocompatibility}, %
 final, 			%
 tracking=true,
 kerning=true,
 spacing=true,
 factor=1100, 	%
 stretch=20, 	%
 shrink=20 		%
]{microtype}

\usepackage{enumitem}

\usepackage{amsmath,amsfonts,bm}

\def\eqref#1{equation~\ref{#1}}

\def\1{\bm{1}}

\def\rve{{\mathbf{e}}}

\def\rvh{{\mathbf{h}}}

\def\rvs{{\mathbf{s}}}

\def\rvx{{\mathbf{x}}}

\def\rvz{{\mathbf{z}}}

\def\va{{\bm{a}}}

\def\vp{{\bm{p}}}

\def\mD{{\bm{D}}}

\def\mF{{\bm{F}}}

\def\mW{{\bm{W}}}
\def\mX{{\bm{X}}}
\def\mY{{\bm{Y}}}

\DeclareMathAlphabet{\mathsfit}{\encodingdefault}{\sfdefault}{m}{sl}
\SetMathAlphabet{\mathsfit}{bold}{\encodingdefault}{\sfdefault}{bx}{n}

\def\gE{{\mathcal{E}}}

\def\gG{{\mathcal{G}}}

\def\gI{{\mathcal{I}}}

\def\gM{{\mathcal{M}}}
\def\gN{{\mathcal{N}}}

\def\gV{{\mathcal{V}}}

\def\sR{{\mathbb{R}}}

\newcommand{\E}{\mathbb{E}}

\usepackage{etoolbox}
\makeatletter
\patchcmd{\@begintheorem}{\textit}{\textbf}{}{}
\makeatother
\newtheorem{thm}{Theorem}[section]

\newtheorem{prop}[thm]{Proposition}
\newtheorem{corr}[thm]{Corollary}

\makeatletter
\newcommand{\algmargin}{\the\ALG@thistlm}
\makeatother

\renewcommand{\algorithmicrequire}{\textbf{Input:}}
\renewcommand{\algorithmicensure}{\textbf{Output:}}

\newcommand{\name}{{InfoMotif}}
\def\BibTeX{{\rm B\kern-.05em{\sc i\kern-.025em b}\kern-.08em
    T\kern-.1667em\lower.7ex\hbox{E}\kern-.125emX}}
    
\makeatletter
\patchcmd{\@maketitle}
  {\addvspace{0.5\baselineskip}\egroup}
  {\addvspace{-1\baselineskip}\egroup}
  {}
  {}
\makeatother

\begin{document}
\renewcommand*{\thefootnote}{\fnsymbol{footnote}}

\title{Beyond Localized Graph Neural Networks: An Attributed Motif Regularization Framework}

\newtheorem{definition}{Definition}

\author{\IEEEauthorblockN{Aravind Sankar$^*$, Junting Wang$^*$, Adit Krishnan, Hari Sundaram}
\IEEEauthorblockA{University of Illinois at Urbana-Champaign, IL, USA\\
\{asankar3, junting3, aditk2, hs1\}@illinois.edu}
}

\stepcounter{footnote}

\maketitle
\stepcounter{footnote}
\footnotetext{Equal contribution}

\begin{abstract}
We present InfoMotif, a new semi-supervised, motif-regularized, learning framework over graphs. We overcome two key limitations of message passing in popular graph neural networks (GNNs): \textit{localization} (a $k$-layer GNN cannot utilize features outside the $k$-hop neighborhood of the labeled training nodes) and \textit{over-smoothed} (structurally indistinguishable) representations.
We propose the concept of \textit{attributed structural roles} of nodes based on their occurrence in different \textit{network motifs}, independent of network proximity.
Two nodes share attributed structural roles if they participate in topologically similar motif instances over co-varying sets of attributes. Further, InfoMotif achieves architecture independence by regularizing the node representations of arbitrary GNNs via \textit{mutual information} maximization.
Our training curriculum 
dynamically prioritizes multiple motifs in the learning process without relying on distributional assumptions in the underlying graph or the learning task.
We integrate three state-of-the-art GNNs in our framework, to show significant gains (3--10\% accuracy)
across six diverse, real-world datasets. We see stronger gains for nodes with \textit{sparse training labels} and \textit{diverse attributes} in local neighborhood structures.
\end{abstract}

\renewcommand{\thefootnote}{\arabic{footnote}}
\setcounter{footnote}{0}

\section{Introduction}

This paper proposes a class of motif-regularized graph neural networks (GNNs); GNNs have emerged as a popular paradigm for semi-supervised learning on graphs due to their ability to learn representations combining topology and attributes.
GNNs are typically formulated as a message passing framework~\cite{gnn_review},
where the representation of a node is computed by a GNN layer aggregating features from its graph neighbors via learnable aggregators.
Long-range dependencies are captured by using $k$ layers to incorporate features from $k$-hop neighborhoods.

\begin{figure}[htbp]
    \centering
    \includegraphics[width=\linewidth, trim={0in 6.1in 3.5in 0in}, clip]{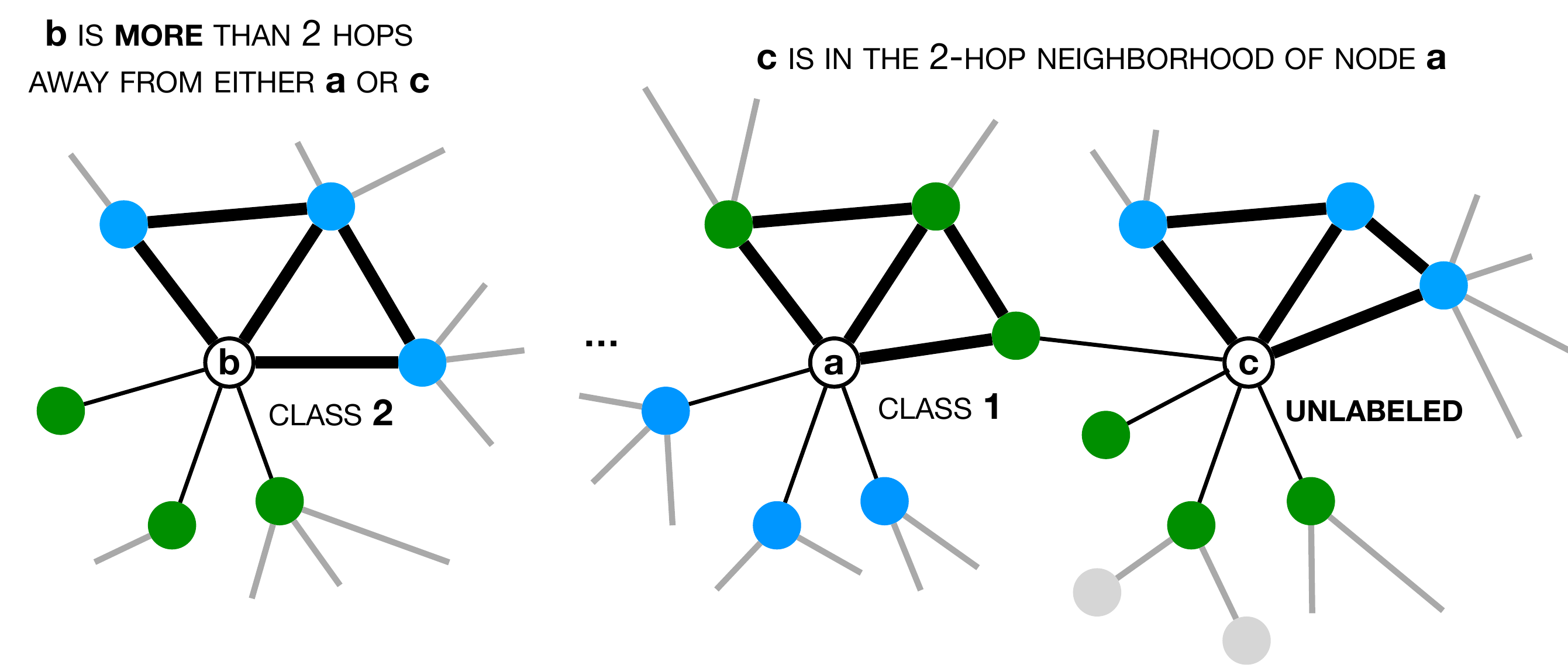}
    \caption{\textbf{Localized message passing limitations: } A stylized example with a 2-layer GNN (colors indicate node attributes). Node $a$ is in the 2-hop range of node $c$. Node $c$ \textbf{does not receive} gradient updates from node $b$ (class 2) since node $b$ is more than 2 hops away. The GNN will likely label node $c$ as class 1. Notice that $c$ is in class 2 as since $c$ and $b$ have identical local structure and attribute co-variation.
    }
    \label{fig:example}
\end{figure}

\textbf{Localized message passing limitations: } 
We illustrate two key limitations of prior $k$-layer GNN architectures: \textit{$k$-hop localized} and \textit{over-smoothed} representations (\Cref{fig:example}).

\begin{enumerate}[leftmargin=*]
    \item GNNs, while highly expressive, are inherently \textit{localized}: a $k$-layer GNN cannot utilize features of nodes that lie outside the $k$-hop neighborhood of the labeled training nodes.  In~\Cref{fig:example}, nodes $a$ and $b$ belong to different classes.
          A 2-layer GNN
          sees unlabeled node $c$ within the aggregation range of $a$ (class 1) and outside the influence of $b$ (class 2 and more than 2 hops away). Thus,
          a GNN will more likely label $c$ with class 1 (than class 2).
          However, in reality, $c$ and $b$ display identical attributes (node color) in the local structure; a localized GNN fails to incorporate this factor.
    \item
          GNNs with multiple layers learn \textit{over-smoothed} node representations by iteratively aggregating neighbor features
          ~\cite{gcn_oversmoothing}.
          In~\Cref{fig:example}, nodes $c$ and $a$ share the same number of neighbors with blue and green attributes; however, green neighbors of node $a$ form triangles, while blue neighbors of node $b$ (and $c$) form triangles.
          Considering \textit{local nodal attribute arrangements}, node $c$ is more similar to $b$ than to $a$.
          The over-smoothing effect in GNNs obscures this attribute co-variation difference when classifying node $c$.
\end{enumerate}

Thus, we require a new learning framework over graphs, to overcome the limitations of message passing in popular GNNs.

One way to overcome these limitations is the paradigm of role discovery~\cite{role_discovery} that identifies nodes with structurally similar neighborhoods.
In contrast to the notion of communities defined by network proximity, structural roles characterize nodes by their local connectivity and subgraph patterns independent of their location in the network~\cite{rossi2019community}; thus, two nodes with similar roles may lie in different parts of the graph.
Prior role-aware models learn similar representations for structurally similar nodes while ignoring nodal attributes~\cite{struc2vec}, \textit{i.e.}, they will assign the same role to nodes $a$ and $b$ in~\Cref{fig:example} with topologically identical local structures; however, nodes $a$ and $b$ differ in their local attribute arrangements (blue vs. green attributes in triangles), and thus belong to different classes.

\textbf{Present Work: } To enable the expressivity to distinguish attributed structures, we propose the concept of
\textit{attributed structural roles} that identify %
structurally similar nodes with co-varying attributes,
independent of network proximity.
We ground structural roles on \textit{network motifs}\footnote{The terms network motif, graphlet, and induced subgraph are used interchangeably in literature}, 
which are higher-order structures crucial to the organization of complex networks~\cite{network_motif}.
We define two nodes as sharing attributed structural roles if they participate in topologically similar motif instances over co-varying sets of attributes. We note that attribute co-variance permits for multiple discrete and continuous attributes, rather than stricter notions such as regular equivalence~\cite{rossi2019community}.

We propose~\name, a GNN architecture-agnostic regularization framework 
that exploits the co-variance of attributes and motif structures.
~\name~learns regularizers based on a set of network motifs, which vary in their task-specific significance.
Specifically, across instances of the same motif (\textit{e.g.}, a triangle structure), we learn discriminative attribute correlations to regularize the underlying GNN node representations;  
this encourages the GNN to learn statistical correspondences between distant nodes that participate in similarly attributed instances of that motif.
We propose a novel training curriculum to integrate multiple motif regularizers while attending to motif types and  skewed motif distributions. Our key contributions:

\begin{itemize}[leftmargin=*]
    \item \textbf{Attributed Structural Role Learning}:
          We propose the novel concept of \textit{attributed structural roles} to regularize GNN models for semi-supervised learning.
          In contrast to prior work that identify structurally similar nodes agnostic to attributes~\cite{struc2vec}, we use the principle of \textit{mutual information maximization} to regularize node representations to capture \textit{attribute correlations} in motif structures. ~\name~unifies the expressive local neighborhood aggregation power of GNNs with the paradigm of structural role discovery.
    \item \textbf{Architecture-agnostic Regularization Framework:}
          To the best of our knowledge,~\name~is the first to address the limitations of localized message passing in GNNs through an architecture-agnostic framework.
          Unlike prior attempts that design new aggregators~\cite{demonet,jknet},
           we achieve architecture independence by modulating the node representations learned by the base GNN, to capture attributed structural roles. 
     We show significant gains over the state-of-the art GNNs.
    \item \textbf{Distribution-agnostic Multi-Motif Curriculum}:
          We propose two learning progress indicators, \textit{task-driven utility} and \textit{distributional novelty}, to integrate multiple motif regularizers within our framework.
          Unlike prior strategies~\cite{transductive, dualgcn} that incorporate regularizers via tunable hyper-parameters, our training curriculum dynamically prioritizes different motifs in the learning process without relying on distributional assumptions on the underlying graph or on the learning task.
\end{itemize}

We integrate three state-of-the-art GNN models in our framework, to show significant gains (3-10\% accuracy) with motif-based regularization on two diverse classes of datasets: \textit{citation} networks that exhibit strong homophily and \textit{air-traffic} networks that depend on structural roles. Our qualitative analysis %
indicates stronger gains for nodes with \textit{sparse training labels} and \textit{diverse attributes} in local neighborhood structures.

We organize the rest of the paper as follows. In Section~\ref{sec:formulation}, we
present the problem formulation, and introduce preliminaries on GNNs and network motifs.
We describe our proposed framework~\name~in Sections~\ref{sec:methods} and~\ref{sec:model_details}, present
experimental results in Section~\ref{sec:experiments}, finally concluding in Section~\ref{sec:conclusion}.

\section{Preliminaries}
\label{sec:formulation}
In this section, we formalize semi-supervised node classification on graphs via Graph Neural Networks and introduce network motifs to regularize the classification.

\subsection{\textbf{Problem Definition}}
\label{sec:prob_defn}
Let $\mathcal{G} = (\mathcal{V}, \mathcal{E})$ be an attributed graph, with nodes $\mathcal{V}$ and edges $\mathcal{E} \in \mathcal{V} \times \mathcal{V}$. Note, $\mathcal{V} = \mathcal{V}_L \cup \mathcal{V}_U$, the sets of labeled ($\mathcal{V}_L$) and unlabeled ($\mathcal{V}_U$) nodes in the graph.
Let $\mathcal{N}(v)$ denote the neighbor set of node $v \in \mathcal{V}$ in $\mathcal{G}$, and $\mathbf{X} \in \mathbb{R}^{|\mathcal{V}| \times F}$ denotes the attribute matrix with rows $\mathbf{x}_v \in \mathbb{R}^F$ for node $v \in \mathcal{V}$.
Each labeled node $v \in \mathcal{V}_L$ belongs to one of $C$ classes, encoded by a one-hot vector $\mathbf{y}_v \in \mathbb{B}^C$ ($\mathbb{B} = \{0,1\}$). Our goal is to predict the labels %
of the unlabeled nodes $v \in \mathcal{V}_U$. This is the familiar transductive learning setup for node classification~\cite{transductive}. %

\subsection{\textbf{Graph Neural Networks}}
\label{sec:base_gnn}
Graph Neural Networks (GNNs) use multiple layers to learn node representations. At each layer $l>0$, where $0$ is the input layer, GNNs compute a representation for node $v$ by aggregating features from its neighborhood, through a learnable aggregator function $f_{\theta, l}$ per layer. Using $k$ layers allows for the $k$-hop neighborhood of a node to influence its representation.

Let $\mathbf{h}_{v,l-1} \in \mathbb{R}^{D}$ denote the representation of node $v$ in layer $l-1$. The $l$-th layer follows a message passing rule:

\begin{equation}
\mathbf{h}_{v,l} = f_{\theta,l} \Big( \mathbf{h}_{v, l-1}, \{  \mathbf{h}_{u, l-1} \} \Big), \quad u \in \mathcal{N}_v
\label{eq:basic_agg}
\end{equation}	

~\Cref{eq:basic_agg} says that the node embedding $\mathbf{h}_{v,l} \in \mathbb{R}^{D}$ for node $v$ at the $l$-th layer is a non-linear aggregation $f_{\theta, l}$ of the embeddings from layer $l-1$ of node $v$ and the embeddings of immediate network neighbors $u \in \mathcal{N}(v)$ of node $v$. The function $f_{\theta, l}$ defines the message passing mechanism at layer $l$
and we can use a variety of aggregator architectures, including graph convolution~\cite{gcn}, graph attention~\cite{gat}, and pooling~\cite{graphsage}. The node representation for $v$ at the input layer is $\mathbf{h}_{v, 0}$  (\textit{i.e.}, $l=0$), where $\mathbf{h}_{v, 0} = \rvx_v$ and $\rvx_v \in \mathbb{R}^{F}$.
We designate the representation of node $v$ at the final GNN layer $\rvh_v \in \sR^{D}$, as its \textbf{base GNN representation}. %
In this work, we use GNNs as a collective term for networks that operate over graphs using localized message passing, as opposed to spectral methods~\cite{spectral} that learn convolutional filters from the entire graph.

\subsection{\textbf{Network Motifs}}
\label{sec:network_motif}

\textit{Network motifs} are a general class of higher-order connectivity patterns, with a history of use in network science~\cite{network_motif,temporal_motif}. A motif has several topologically equivalent appearances in the network %
called \textit{motif instances}. Prior work~\cite{graphlet_counts, subgraph_counting_survey} shows how to efficiently compute motif instances for large graphs.

\begin{definition}[Network Motif]
A network motif $M_t = (\gV_t, \gE_t)$ is a connected, induced subgraph consisting of a subset $\mathcal{V}_t \subset \mathcal{V}$ and $\mathcal{E}_t = \{ e \in \mathcal{E} \mid e = (u, v),  u, v \in \mathcal{V}_t \}$. Let $k_t$ be the number of nodes in $M_t$; that is, $k_t=|\mathcal{V}_t|$. We assume that a graph has a set of unique associated motifs $\mathcal{M} = \{ M_1, \dots, M_T\}$.
\end{definition}

\begin{definition}[Motif Instance] Let $I_t$ be an induced subgraph of $\gG$. We define $I_t$ to be a motif instance of $M_t$ if $I_t$ is isomorphic to $M_t$. A motif $M_t$ can have several motif instances in $\gG$. While each such motif instance has a unique node set, two motif instances can share nodes. We denote the set of unique instances of $M_t$ in $\mathcal{G}$ that contain node $v$ as $\mathcal{I}_v (M_t)$.
\end{definition}

\begin{figure}[htbp]
    \centering
    \includegraphics[width=0.9\linewidth]{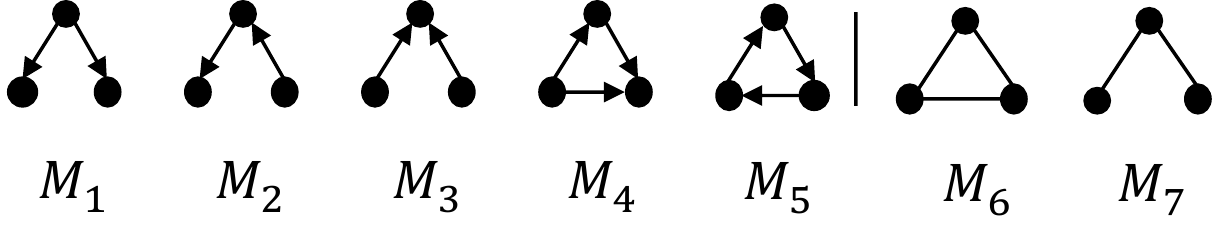}
    \caption{Topologically distinct, directed ($M_1$ to $M_5$) and undirected ($M_6$ to $M_7$) 3-node, connected, network motifs}
    \label{fig:motifs}
\end{figure}

In this work, we consider 3-node connected network motifs, \textit{e.g.},~\Cref{fig:motifs} shows all 3-node, topologically distinct, directed (\textit{e.g.}, citations) and undirected, connected network motifs.

\subsection{\textbf{Regularization}}
\label{sub:Regularization}
We plan to use these local structural properties (\textit{i.e.}, network motifs) to regularize the graph neural model during training. Typically, we train GNNs by minimizing the cross-entropy loss $L_B$, between model predictions $\mathbf{\hat{y}}_v \in \mathbb{R}^{C}$ and ground-truth labels $\mathbf{y}_v \in \mathbb{B}^{C}$ of training nodes in $v \in \mathcal{V}_L$, defined by:
\begin{equation}
    L_B = - \sum\limits_{v \in \mathcal{V}_L} \sum\limits_{c=1}^C y_{v, c} \log \hat{y}_{v, c}
    \label{eqn:base_supervised_loss}
\end{equation}
where, the $c$-th index of the one-hot vector $\hat{y}_{v,c}$ refers to the probability that $v$ belongs to the true class $c$. 
Notice that the loss $L_B$ is agnostic to any local structural properties (\textit{e.g.}, mixing patterns in social networks~\cite{multiscale}) that may be indicative of the true node class. Thus, we develop a modified loss $L'_B = L_B + \lambda L_R$, where $L_R$ is the regularization loss that incorporates attributed motif structure and $\lambda$ is a constant. %
Our goal is to design $L_R$ to overcome the two limitations of message-passing models: localized and over-smoothed node representations.

\begin{table}[t]
    \centering
    \begin{tabular}{@{}rl@{}}
        \toprule
        Symbol                & Description                                                               \\
        \midrule
        $\mathcal{M}$         & Set $\{M_1, \dots, M_T\}$ of $T$ network motifs                           \\
        $\mathcal{I}_v (M_t)$ & Set of instances of motif $M_t$ in $\mathcal{G}$ that contain node $v$    \\
        $\mathbf{h}_{v, l}$   & Representation of node $v$ at layer $l$ of GNN                            \\
        $\mathbf{h}_v$        & Base GNN representation of node $v$ (final layer)                         \\
        $\mathbf{h}^{t}_v$    & Motif-gated representation of node $v$ for motif $M_t$                    \\
        $\mathbf{e}_{v, I_t}$ & Instance-specific representation of $v$ in $I_t \in \mathcal{I}_v (M_t) $ \\
        $\mathbf{s}_{v, t}$   & Motif-level representation of node $v$ for motif $M_t$                    \\
        $\mathbf{z}_v$        & Final Representation of node $v$                                          \\
        $\alpha_{vt}$         & Task-specific importance of motif $M_t$ to node $v$                       \\
        $\beta_v$             & Novelty score for training node $v \in \gV_L$                             \\
        \bottomrule
    \end{tabular}
    \caption{Notation}
    \label{tab:notations}
\end{table}

\section{InfoMotif Framework}
\label{sec:methods}
In this section, we first discuss the structural properties of GNNs to motivate the notion of attributed structural roles.
In section~\ref{sec:single_motif}, we present our motif-based mutual information maximization framework~\name~to regularize GNNs based on a single motif. %
Finally, in section~\ref{sec:multi_motif}, we introduce our overall framework with a novel multi-motif training curriculum.

\subsection{\textbf{Attributed Structural Role Learning}}
\label{sec:role_learning}
A $k$-layer GNN computes a localized representation $\mathbf{h}_{v, k}$ for each node $v$ that incorporates information from its $k$-hop neighborhood, denoted by $\mathcal{N}_k (v)$. %
For a node set $S \subseteq \gV$, let $\mathcal{N}_k(S) = \bigcup_{v \in S} \gN_k (v) $ define its $k$-hop neighborhood, and $\mX (S)$ denote its set of input node features. 
Let $\mY (\gV_L)$ comprise the training labels of nodes in the labeled set $\gV_L$.
For a $k$-layer GNN trained on 
$\gV_L$ using loss $L_B$ (\Cref{eqn:base_supervised_loss}), let 
$\Theta^* = \{ \Theta_{1}, \dots, \Theta_{k} \}$ be the optimal parameters computed by its training algorithm.
Now, we have the following proposition.
\begin{prop}
$\Theta^{*}$ is a function of $\mX (\mathcal{N}_k(\gV_L)), \mY (\gV_L)$ and changes in inputs $\mX (\gV \setminus \mathcal{N}_k(\gV_L))$ will not affect $\Theta^{*}$.
\label{lem:lemma}
\end{prop}

\textit{Proof Sketch.}
By an induction argument, the loss $L_B$ can be written as $g(\Theta_{1}, \dots, \Theta_{k}, \mY (\gV_L), \mX (\mathcal{N}_k(\gV_L))$ for some function $g(\cdot)$.
Thus, when the GNN is trained on $L_B$ using gradient updates, $\Theta^{*}$ must be independent of $\mX (\gV \setminus \mathcal{N}_k(\gV_L))$.

Note that addition of a standard regularization term (\textit{e.g.}, $L_1$ or $L_2$) only impacts $\{\Theta_{1}, \dots, \Theta_{k} \}$; the overall loss still remains independent of $\gV \setminus \mathcal{N}_k(\gV_L)$, satisfying proposition~\ref{lem:lemma}.

Thus, the optimal parameters of a $k$-layer GNN are only affected by node features in the $k$-hop neighborhood $\gN_k(\gV_L)$ of the labeled set $\gV_L$, \textit{i.e.}, the features and connectivities of nodes in $\gV \setminus \mathcal{N}_k(\gV_L)$ are ignored in the training process.

Let the $k$-hop neighborhood of class $c$ be $\gN_k ( \gV_L (c) )$ where $\gV_L (c) = \{ v \in \gV_L : y_{vc} = 1\}$ is the set of nodes labeled with class $c$.
Let $L_B(c)$
be the supervised loss term specific to class $c$. 
Now, the corollary directly follows from proposition~\ref{lem:lemma}:
\begin{corr}
If node $v \not \in \gN_k ( \gV_L (c) ) $, the $k$-hop neighborhood of class $c$, then the loss $L_B (c)$ is independent of $v$.
\label{lem:corr}
\end{corr}
\vspace{-10pt}

The above corollary states that gradient updates from the supervised loss $L_B (c)$ for class $c$ cannot reach nodes that lie outside the $k$-hop neighborhood of class $c$, \textit{i.e.}, $\gN_k ( \gV_L (c) )$.

To illustrate its implications, we revisit~\Cref{fig:example}. Since node $c$ lies beyond the $2$-hop neighborhood of node $b$, node $c$ does not affect the training loss at node $b$ (which belongs to class 2). 
Thus, despite nodes $c$ and $b$ having identical co-variation of attributes and structure (blue
neighbors form triangles), node $c$ does not influence the training loss for all nodes with class 2.

\begin{figure*}[t]
    \centering
    \includegraphics[width=\linewidth]{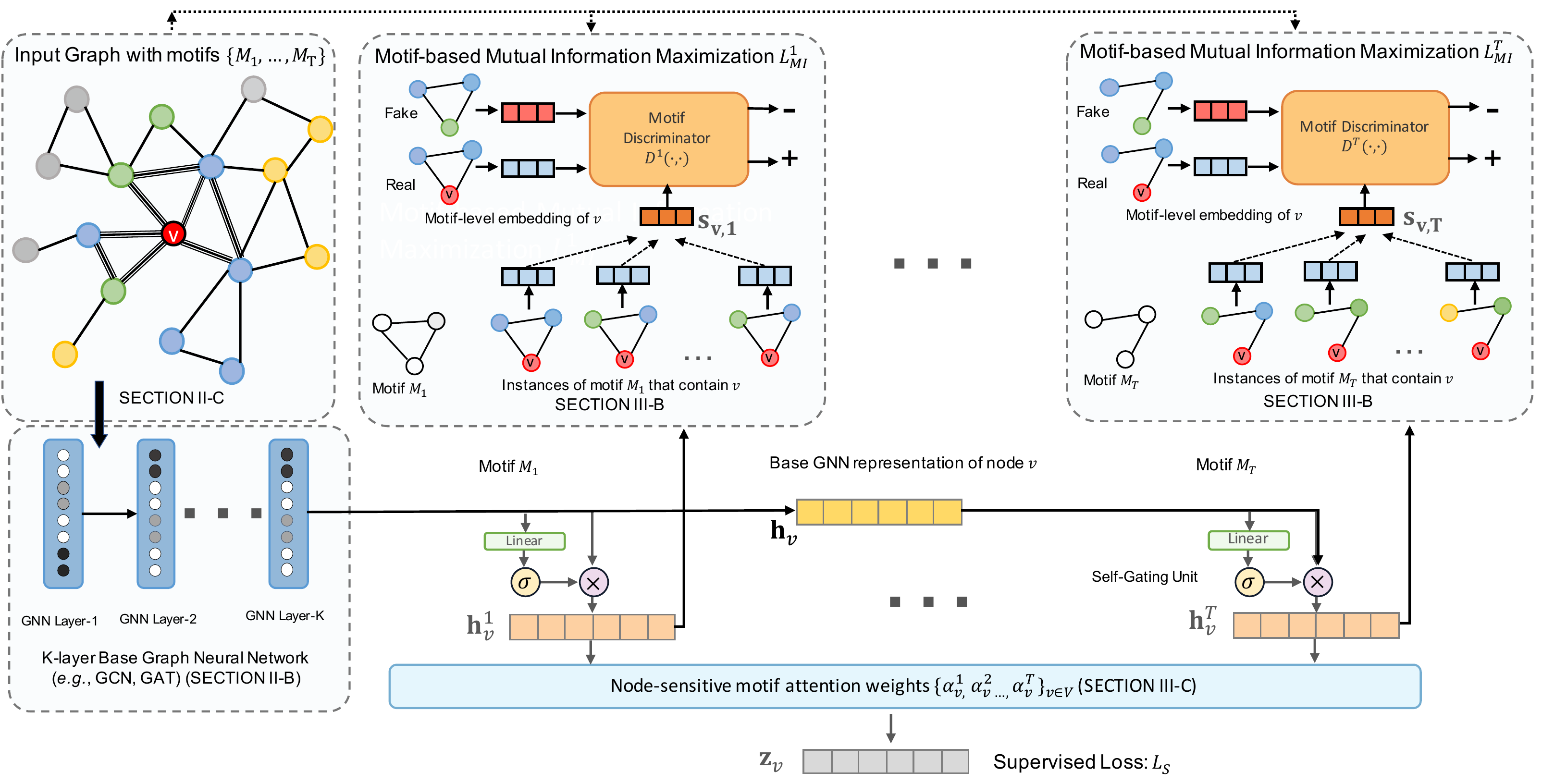}
    \caption{Architecture diagram of~\name~depicting the model components: base GNN $f_{\theta, l}$ with $k$ layers (bottom left), motif-based mutual information maximizing regularizers $L^{t}_{MI}$ (top right), and attention module to compute final node representations $\rvz_v$ (bottom right).
    Instances of motif $M_1$ are shown in the graph (top left) with textured lines and colors indicate node attributes.}
    \label{fig:framework}
    \vspace{-10pt}
\end{figure*}

\subsection{\textbf{Single Motif Regularization}}
\label{sec:single_motif}
In this section, we introduce~\name, a framework to regularize node representations of the base GNN by exploiting the co-variance of node attributes and motif structures. We define \textit{attributed structural roles} by assigning the same role to nodes that participate in motif instances over \textit{co-varying} sets of attributes. In contrast,  prior role-aware models~\cite{struc2vec} discover structurally similar nodes \textit{agnostic} to attributes.

Now, we describe our regularization strategy to learn attribute co-variance for a single motif. In the next section, we extend these arguments to handle multiple motifs.

\subsubsection*{\textbf{Motif-based Mutual Information}}
We first consider a single network motif type $M_t \in \mathcal{M}$ and a specific node $v \in \gV$ to learn attribute co-variance across instances $\mathcal{I}_v (M_t)$ that contain $v$ in the graph.
To learn attributed structural roles, it is necessary to \textit{contrast} the attributed instances of motif $M_t$ against attributed node combinations that are not present in any instances of $M_t$.

We maximize the motif-based \textit{mutual information} (MI) between a \textit{motif-level} representation of $v$ and corresponding \textit{instance-specific} representations centered at $v$.
By introducing motif-based MI maximization as a regularizer, the GNN is encouraged to learn \textit{discriminative} statistical correspondences between nodes that participate in instances of the same motif.

We first adapt the base GNN representation $\mathbf{h}_v$ (see~\Cref{sec:base_gnn}), specific to motif $M_t$ through a \textit{motif gating} function $f^{t}_{\textsc{gate}} :  \mathbb{R}^D \mapsto  \mathbb{R}^D$ resulting in a gated embedding $\mathbf{h}_v^{t}$.
Then, we introduce a \textit{motif instance encoder} $f^{t}_{\textsc{enc}}:  \mathbb{R}^D \times \mathbb{R}^{(k_t \times D)} : \mapsto \mathbb{R}^D $ to compute the instance-specific representation $\mathbf{e}_{v, I_t} \in \mathbb{R}^D$ of node $v$ conditioned on other co-occurring nodes in instance $I_t \in \mathcal{I}_v (M_t)$.
Finally, the motif-level representation $\mathbf{s}_{v, t} \in \mathbb{R}^D$ of node $v$ summarizes the set of instance-specific representations $\{ \mathbf{e}_{v, I_t} \}_{I_t \in \mathcal{I}_v (M_t)}$ through a permutation-invariant   \textit{motif readout} function $f^{t}_{\textsc{read}} (\cdot)$, \textit{e.g.}, averaging or pooling functions.

For each node $v \in \mathcal{V}$, we maximize motif-based mutual information $L^{t}_{MI}$ between its instance-specific representations $\{ \rve_{v, I_t} \}_{I_t \in \mathcal{I}_v (M_t)}$ and motif-level representation $\rvs_{v, t}$, by defining $I_{\psi^t}$ as a mutual information estimator for motif $M_t$ that is \textit{shared} across all nodes.
The resulting objective is given by:

\begin{equation}
L^{t}_{MI} (\theta, \phi^t, \psi_t) = \frac{1}{|\gV|} \sum\limits_{v \in \gV} \sum\limits_{I_t \in \gI_v(M_t)} I_{ \psi_t} ( \rve_{v, I_t} ; \rvs_{v, t} )
\end{equation}
\normalsize

where 
$\theta$ and $\phi^t$ denote the parameters of the layers $\{ f_{\theta, l}\}_{l=1}^k$, and motif-specific transforms $\{f^{t}_{\textsc{gate}}, f^{t}_{\textsc{enc}}, f^{t}_{\textsc{read}} \}$ respectively.
By maximizing MI across all instances of motif $M_t$ in the graph through a shared MI estimator $I^{t}_{\psi}$, we enable the GNN to learn correspondences between a pair of potentially distant nodes that participate in instances of motif $M_t$.

\subsubsection*{\textbf{Mutual Information Maximization}}
Following neural MI estimation methods~\cite{mine, dim}, 
we model the estimator $I_{\psi^t}$ as a \textit{discriminator} network that learns a decision boundary to accurately distinguish between \textit{positive} samples drawn from the joint distribution and \textit{negative} samples drawn from the product of marginals.
We train a \textit{constrastive} discriminator network $\mD^t_{\psi} : \sR^D \times \sR^D \mapsto \sR^{+} $, where $\mD_{\psi}^{t}(\rve_{v, I_t} , \rvs_{v, t})$ denotes the probability score assigned to this instance-motif pair.
The positive samples $(\rve_{v, I_t}, \rvs_{v, t})$ for $\mD_{\psi}^{t}$ are the representations of observed instances $I_t \in \gI_v(M_t)$ of motif $M_t$ paired with the motif-level representation $\rvs_{v,t}$. 
The negative samples  $(\rve_{v, \widetilde{I}_t}, \rvs_{v,t})$ are derived by pairing $\rvs_{v, t}$ with the representations of negative instances $\widetilde{I}_t$ sampled from a distribution $P_{\gN} (\widetilde{I}_t | M_t)$.
The discriminator  $\mD^t_{\psi}$ is trained on a noise-contrastive objective $L^{t}_{MI}$
between samples from the joint (positive pairs), and the product of marginals (negative pairs), which is defined as: %

{\footnotesize
\begin{align}
    L^{t}_{MI} = \frac{1}{|\gV|} \sum\limits_{v \in \gV} L^{t}_{MI} (v)  = & - \frac{1}{2Q |\gV|}  %
    \sum\limits_{v \in \gV}  %
       \sum\limits_{i=1}^Q \Big[  \E_{I_t}  \log  \mD_{\psi}^{t} (\rve_{v, I_t} , \rvs_{v, t})  \nonumber \\   & +
     \E_{\widetilde{I}_t} \log (1 - \mD_{\psi}^{t} (\rve_{v, \widetilde{I}_t} , \rvs_{v, t}) ) \Big]
    \label{eqn:mi_loss}
\end{align}
}%

where $Q$ is the number of observed motif instances sampled per node. 
This objective maximizes MI between $\rvs_{v ,t}$ and  $\{ \mathbf{e}_{v, I_t} \}_{I_t \in \mathcal{I}_v (M_t)}$ based on the Jensen-Shannon Divergence between their joint distribution and  product of marginals~\cite{dgi}.

We design the negative sampling distribution $P_{\gN} (\widetilde{I}_t | M_t)$ to learn attribute co-variance in instances of motif $M_t$.
For each positive instance $I_t$, the generated negative instance $\widetilde{I}_t$ is topologically equivalent but contains attributes that do not occur in instances of $M_t$ in $\gG$.
By contrasting the observed instances of $M_t$ against fake instances with perturbed attributes, $\mD_{\psi}^{t}$ learns attributed structural roles with respect to motif $M_t$.

\subsection{\textbf{Multi-Motif Regularization Framework}}
\label{sec:multi_motif}
Now, we extend our framework for any graph that includes a set of motifs $\gM = \{ M_1, \dots, M_T\}$. 
A typical way to include regularizers (\Cref{eqn:mi_loss}) from multiple motifs is given by:
\begin{equation}
L^{'} = L_B + \lambda L^{'}_{MI} = L_B + \lambda \cdot \frac{1}{T} \sum\limits_{t=1}^T L^{t}_{MI}
\label{eqn:base_reg}
\end{equation}

where $\lambda$ is a tunable hyper-parameter to balance the supervised task loss $L_B$ and motif regularizers.
Intuitively, each motif $M_t \in \mathcal{M}$ is a connectivity pattern that can be viewed as defining one kind of structural role, \textit{e.g.}, bridge nodes.
Each motif has a different significance towards the learning task.
Thus, a multi-motif framework should automatically identify the significance of different motifs without manual hand tuning.

In addition, real-world networks exhibit heavy-tailed degree and community distributions~\cite{tail}, which manifest as \textit{skewed} (imbalanced) motif occurrences among nodes as well as across motif types. 
This further complicates the learning process of incorporating multiple motifs as regularizers.
We identify three key aspects \textit{task}, \textit{node}, and \textit{skew} for a multi-motif framework:
\begin{itemize}[leftmargin=*]
\item \textbf{Task}: Distinguish the significance of different motifs to compute representations conditioned on the learning task.
\item \textbf{Node}: Expressive power to control the extent of regularization exerted by each motif at a node-level granularity.
\item \textbf{Skew}: Adapt to varying levels of motif occurrence skew without any distributional assumptions on the input graph.
\end{itemize}

To address these objectives, we first describe our approach to compute final node representations conditioned on multiple motifs, followed by two novel online reweighting strategies. %

\subsubsection*{\textbf{Task-driven Representations}}
The base GNN is trained by a supervised task loss $L_B$ (\Cref{eqn:base_supervised_loss}) over the labeled node set $\gV_L$.
We instead aggregate the set of motif-gated representations ($\rvh_v^t$ for motif $M_t \in \gM$), to compute the final representation $\rvz_v \in \sR^{D}$ for node $v$.
We learn attention weights $\alpha_{vt}$ to characterize the task-driven importance of motif $M_t$ to node $v$ and compute $\rvz_v$ through a weighted average, given by:
\begin{equation}
\rvz_v = \sum\limits_{t=1}^{T} \alpha_{vt} \rvh^{t}_v  \hspace{10 pt} 
\alpha_{vt} = \frac{\exp \big(\vp \cdot  \rvh^{t}_v  \big) }{\sum\limits_{t'=1}^T \exp \big(\vp  \cdot \rvh^{t'}_{v} \big)} 
\label{eqn:motif_attention}
\end{equation}
\normalsize

where $\vp \in \sR^D$ defines the attention function and is learned by optimizing the final representations $\{ \rvz_v \}_{v \in \gV_L}$ of labeled nodes $\gV_L$ using the supervised loss $L_B$ (\Cref{eqn:base_supervised_loss}). The final representation $\rvz_v$ of each node $v \in \gV$ is used for classification.

\subsubsection*{\textbf{Node-sensitive Motif Regularization}}

Instead of using static uniform weights to incorporate motif regularizers (\Cref{eqn:base_reg}), we contextually weight the contributions of different motif regularization terms (\Cref{eqn:mi_loss}) at a node-level granularity through the attention weights $\alpha_{vt}$ of motif $M_t$ for node $v$.

\begin{equation}
 L_{MI} = \frac{1}{nT}\sum\limits_{t=1}^T \sum\limits_{v \in \gV} \alpha_{vt} L^t_{MI} (v) 
 \label{eqn:full_mi_loss}
\end{equation}

\normalsize
The loss $L_{MI}$ varies the extent of regularization per node in proportion to the task-specific importance $\alpha_{vt}$ of motif $M_t$ to node $v$.
Notice that while the attention function is learned by training the final representations $\rvz_v$ of labeled nodes $v \in \gV_L$ on the supervised loss $L_B$, the motif-regularization loss $L_{MI}$ (which operates on all nodes) re-weights each motif loss term per node with the estimated attention weights.

\renewcommand{\algorithmicrequire}{\textbf{Input:}}
\renewcommand{\algorithmicensure}{\textbf{Output:}}
\renewcommand{\algorithmiccomment}[1]{$\triangleright$~#1}

\begin{algorithm}[t]
\caption{The framework of~\name-GNN.}
\begin{algorithmic}[1]
\REQUIRE Graph $\gG$, Labeled node set $\gV_L$, Base GNN $\{f_{\theta, l} \}_{l=1}^k$
\ENSURE Motif-regularized embedding $\rvz_v$ for each node $v \in \gV$
\vspace{-9pt}
\STATE Initialize sample novelty weights $\beta_v = 1 \; \forall \; v \in \gV_L$
\WHILE{\textit{not converged}}
\STATE \COMMENT {\textbf{\textit{Supervised loss over labeled node set $\gV_L$}}}
\FOR{each batch of nodes $\gV_B \subseteq \gV_L$}
\STATE Fix sample weights $\{ \beta_v\}_{v \in \gV_B}$ and optimize $L_S$ on $\gV_B$ using mini-batch gradient descent (Equation~\ref{eqn:supervised_loss}).
\ENDFOR
\STATE Compute motif attention weights $\{ \bm{\alpha_v}\}_{v \in \gV}$ (Equation~\ref{eqn:motif_attention}).
\vspace{-9pt}
\STATE \COMMENT {\textbf{\textit{Motif-based InfoMax loss over entire node set $\gV$}}}
\FOR{each batch of nodes $\gV_B \subseteq \gV$}
\STATE Fix motif weights $\{ \bm{\alpha}_v\}_{v \in \gV}$ and optimize $L_{MI}$ on $\gV_B$ using mini-batch gradient descent (Equation~\ref{eqn:full_mi_loss})
\ENDFOR
\STATE Compute sample weights $\{\beta_v\}_{v \in \gV_L}$ (Equation~\ref{eqn:novelty}).
\ENDWHILE
\STATE Compute $\rvz_v \in \sR^D \; \forall \; v \in \gV$ (Equation~\ref{eqn:motif_attention})
\end{algorithmic}
\label{alg:opt}
\end{algorithm}
\subsubsection*{\textbf{Skew-aware Sample Weighting}}
Prior work in curriculum and meta learning has shown the importance of re-weighting training examples to overcome training set biases~\cite{reweight}.
In particular, re-weighting strategies that emphasize harder examples are effective at handling imbalanced data distributions~\cite{imbalance}.
We propose a \textit{novelty-driven} re-weighting strategy to handle skew in motif occurrences across nodes and motif types.

The novelty $\beta_v$ of node $v$ is a function of its motif distribution, \textit{i.e.}, novel nodes contain uncommon motif types in their neighborhood, which in turn reflects in their attention weight distribution over motifs.
Let $\bm{\alpha}_v \in \sR^{T}$ denote the vector of attention weights for a labeled node $v$ over the motif set $\gM$.
Now, the novelty $\beta_v$ of node $v$ is high
if its motif distribution $\bm{\alpha}_v$ significantly diverges from those of other nodes.
We quantify $\beta_v$ by the deviation (measured by euclidean distance) of $\bm{\alpha}_v$ from the mean motif distribution of labeled nodes  $v \in \gV_L$. %
\begin{equation}
 \beta_v = \frac{exp ( \left\lVert \bm{\alpha}_v - \bm{\mu} \right\rVert^2 ) }{ \sum\limits_{u \in \gV_L} \exp (\left\lVert \bm{\alpha}_u - \bm{\mu} \right\rVert^2) }  \hspace{10pt} \bm{\mu} = \frac{1}{|\gV_L|} \sum\limits_{v \in \gV_L} \bm{\alpha}_v
 \label{eqn:novelty}
\end{equation}
\normalsize
The novelty scores are normalized over $\gV_L$ using a softmax function, to give non-negative sample weights $0 < \beta_v \leq 1$. 
We now define the novelty-weighted supervised loss $L_S$ as:
\begin{equation}
    L_S = - \sum\limits_{v \in \mathcal{V}_L} \beta_v \sum\limits_{c=1}^C y_{vc} \log \hat{y}_{vc}
    \label{eqn:supervised_loss}
\end{equation}
\normalsize
In contrast to the original supervised loss $L_B$ (\Cref{eqn:base_supervised_loss}), the re-weighted objective $L_S$ induces a novelty-weighted training curriculum that  progressively focuses on harder samples.

\subsubsection*{\textbf{Model Training}}
The overall objective of~\name~is composed of two terms, the re-weighted supervised loss $L_S$ (\Cref{eqn:supervised_loss}), and motif regularizers (\Cref{eqn:full_mi_loss}), given by:
\begin{equation}
L = L_S + \lambda L_{MI}
\label{eqn:combined_loss}
\end{equation}
\normalsize

In practice, we optimize $L_S$ and $L_{MI}$ alternatively at each training epoch, which removes the need to tune balance hyper-parameter $\lambda$.~\Cref{alg:opt} summarizes the training procedure.

\subsubsection*{\textbf{Complexity Analysis}}
On the whole, the complexity of our model is $O(\mF)+ O(nTQD + nTD^2) $ where $O(\mF)$ is the base GNN complexity, $T$ is the number of motifs, $Q$ is sampled instance count per motif, and $D$ the latent space dimensionality. 
Since $T \ll n$ and $Q \ll n$, the added complexity of our framework scales linearly with respect to the number of nodes.

\section{Model Details}
\label{sec:model_details}
We now discuss the architectural details of our framework: motif instance encoder, gating, readout, and discriminator.

\subsection{\textbf{Motif Gating}}
We design a pre-filter with \textit{self-gating units} (SGUs) to regulate information flow from the base GNN embedding $\mathbf{h}_v$ to the motif-based regularizer.
The SGU  $f^t_{\textsc{gate}} (\cdot)$ for motif $M_t$ learns a non-linear gate to modulate the input at a feature-wise granularity through dimension re-weighting, defined by:
\begin{equation}
    \mathbf{h}^{t}_v = f^t_{\textsc{gate}} ( \mathbf{h}_v) = \mathbf{h}_v \odot \sigma (\mathbf{W}_g^t \mathbf{h}_v  + \mathbf{b}_g^t)
    \label{eqn:gating}
\end{equation}
\normalsize
where $\mathbf{W}^t \in \mathbb{R}^{D \times D}, \mathbf{b}^t \in \mathbb{R}^{D}$ are learned parameters, $\odot$ denotes the element-wise product operation, and $\sigma$ is the sigmoid non-linearity. The self-gating mechanism effectively serves as a multiplicative skip-connection~\cite{glu} that facilitates gradient flow from the motif-based regularizer to the GNN.

\subsection{\textbf{Motif Instance Encoder}} The encoder $f_{\textsc{enc}} (\cdot)$ computes the instance-specific representation $\rve_{v, I_t}$ for node $v$ conditioned on the gated representations $\{ \rvh^t_u\}_{u \in I_t}$ of the nodes in instance $I_t$.
We apply self-attentions~\cite{self_attention} to compute a weighted average of the gated node representations $\{ \rvh^t_u\}_{u \in I_t}$ in $I_t$.
Specifically, $f_{\textsc{enc}}$ attends over each node $u \in I_t$ to compute attention weight $\alpha_{u}$ by comparing its gated representation $\rvh^t_u$ with that of node $v$, $\rvh^t_v$.
\begin{equation}
\rve_{v, I_t} = \sum\limits_{u \in I_t} \alpha_u \rvh^{t}_u  \hspace{10 pt} \alpha_u = \frac{\exp \big(\va^{t} \cdot [ \rvh^{t}_u || \rvh^{t}_v ] \big) }{\sum\limits_{u^{'} \in I_t} \exp \big(\va^t  \cdot [ \rvh^{t}_{u^{'}} || \rvh^{t}_v] \big)}
\label{eqn:encoder}
\end{equation}
\normalsize
where $\va^{t} \in \sR^{2D}$ is a weight vector parameterizing the attention function and $||$ denotes concatenation. We empirically find the self-attentional encoder to outperform other pooling alternatives.

\subsection{\textbf{Motif Readout}}
The readout function $f^t_{\textsc{read}} (\cdot)$ summarizes the set of instance-specific representations $\{\rve_{v, I_t}\}_{I_t \in \gI_v (M_t)}$ to compute the motif-level representation $\rvs_{v ,t}$.
We use a simple averaging of instance-specific representations to define $f^t_{\textsc{read}} (\cdot)$ as follows:
\begin{equation*}
   \rvs_{v ,t} = f^t_{\textsc{read}} \Big(\{\rve_{v, I_t}\}_{I_t \in \gI_v (M_t)} \Big) = \sigma \Big( \sum\limits_{I_t \in \gI_v (M_t) } \frac{\rve_{v, I_t}}{|\gI_v (M_t)|} \Big)
   \label{eqn:readout}
\end{equation*}
 \normalsize
 
where $\sigma$ is the sigmoid non-linearity.
We adopt batch-wise training with motif instance sampling ($\sim$ 20 per node) to compute $\rvs_{v ,t}$. 
Sophisticated readout architectures~\cite{set2vec} are more likely necessary to handle larger sample sizes.

\subsection{\textbf{Motif Discriminator}}
The discriminator $D^{t}_{\psi}$ learns a motif-specific scoring function to assign higher likelihoods to observed instance-motif pairs relative to negative examples.
Similar to prior work~\cite{dgi, groupim}, we use a bilinear scoring function defined by:
\begin{equation}
    \mD^{t}_{\psi} (\rve_{v, I_t}, \rvs^{t}_v) = \sigma (\rve_{v, I_t} \cdot \mW_d^{t} \rvs^t_v)
\end{equation}
\normalsize
where $\mW_d^{t} \in \sR^{D \times D} $ is a trainable scoring matrix and $\sigma$ is the sigmoid non-linearity to convert raw scores into probabilities of $(\rve_{v, I_t}, \rvs^{t}_v)$ being a positive example for motif $M_t$.

\section{Experiments}
\label{sec:experiments}
\newcolumntype{K}[1]{>{\centering\arraybackslash}p{#1}}
\newcolumntype{R}[1]{>{\raggedright\arraybackslash}p{#1}}

 \begin{table}[t]
 \centering
 \begin{tabular}{@{}p{0.17\linewidth}R{0.09\linewidth}R{0.09\linewidth}R{0.09\linewidth}R{0.09\linewidth}R{0.09\linewidth}R{0.09\linewidth}@{}}
 \toprule
 \multirow{2}{*}
 & \multicolumn{3}{c}{ \textbf{Citation Networks}}  & \multicolumn{3}{c}{ \textbf{Air-Traffic Networks}} \\
 \cmidrule(lr){2-4} \cmidrule(lr){5-7}

 \textbf{Dataset} &  \textbf{Cora} & \textbf{Citeseer} & \textbf{Pubmed}  & \textbf{Brazil} & \textbf{Europe} & \textbf{USA} \\ 
 \midrule
 \textbf{\# Nodes} &  2,485 & 2,110 & 19,717 &  131 & 399 & 1,190\\ 
 \textbf{\# Edges} & 5,069 & 3,668 & 44,324  & 1,038 & 5,995 &13,599\\
 \textbf{\# Attributes} & 1,433  & 3,703 & 500 & - & - & -\\
 \textbf{\# Classes} & 7 & 6 & 3 & 4 & 4 & 4 \\
 \bottomrule
 \end{tabular}
 \caption{Dataset statistics of three benchmark citation~\cite{citation} and air-traffic~\cite{struc2vec} networks.
 Ground-truth classes in citation networks exhibit attribute homophily; ground-truth classes in flight networks indicate node structural roles.}
\label{tab:stats}
\end{table}

We present extensive quantitative and qualitative analyses on multiple diverse datasets.
We first introduce datasets, baselines, and experimental setup (Section~\ref{sec:datasets}, \ref{sec:baselines},~\ref{sec:setup}, and~\ref{sec:results}), followed by node classification results in Section~\ref{sec:results}~by integrating three GNN models in our framework.
In Section~\ref{sec:analysis}, we present a qualitative analysis to analyze the impact of  \textit{label sparsity and attribute diversity in local neighborhoods}.
We then conduct an ablation study to understand our gains over the base GNN models in Section~\ref{sec:ablation}, analyze parameter sensitivity in Section~\ref{sec:sensitivity} and model efficiency in Section~\ref{sec:efficiency},
Finally, we discuss limitations and future directions in Section~\ref{sec:discussion}.

\newcommand*{\factor}{0.079}
\newcommand*{\factortraining}{0.007}
\begin{table*}[t]
\centering
\noindent\setlength\tabcolsep{2.9pt}
\begin{tabular}{@{}p{0.13\linewidth}@{\hspace{0pt}}
K{\factortraining\linewidth}K{\factortraining\linewidth}@{\hspace{0pt}}
K{\factor\linewidth}K{\factor\linewidth}K{\factor\linewidth}@{\hspace{10pt}}
K{\factor\linewidth}K{\factor\linewidth}K{\factor\linewidth}@{\hspace{10pt}}
K{\factor\linewidth}K{\factor\linewidth}K{\factor\linewidth}@{}} \\
\toprule
\multirow{1}{*}{\textbf{}} &  \multicolumn{2}{c}{\textbf{Data}} & \multicolumn{3}{c}{\textbf{Cora}}  & \multicolumn{3}{c}{\textbf{Citeseer}} & \multicolumn{3}{c}{\textbf{PubMed}} \\
\cmidrule{2-3} \cmidrule(lr){4-6} \cmidrule(lr){7-9} \cmidrule(lr){10-12}
\multirow{1}{*}{\textbf{Training Ratio}} & \textbf{X} & \textbf{Y} & \textbf{20\%} & \textbf{40\%} & \textbf{60\%} & \textbf{20\%} & \textbf{40\%} & \textbf{60\%} & \textbf{20\%} & \textbf{40\%}  & \textbf{60\%} \\
\midrule
\multicolumn{12}{c}{\textsc{Proximity-based Graph Embedding Methods}} \\
\midrule[0pt]
\textbf{{Node2Vec~\cite{node2vec}} } &  &    
& 75.7 $\pm$ 0.5& 	76.1 $\pm$ 0.5& 	77.6 $\pm$ 0.5& 	
68.1 $\pm$ 0.5	& 69.1 $\pm$ 0.6& 	69.2 $\pm$ 0.4& 	
80.1 $\pm$ 0.6& 	80.2 $\pm$ 0.6& 	80.4 $\pm$ 0.6    \\
\textbf{Motif2Vec~\cite{motif2vec}}  & &   
& 79.0 $\pm$ 0.4& 	79.2 $\pm$ 0.4& 	79.8 $\pm$ 0.5& 	
66.6 $\pm$ 0.4& 	67.1 $\pm$ 0.3& 	68.8 $\pm$ 0.5& 	
79.8 $\pm$ 0.2& 	79.8 $\pm$ 0.4& 	79.9 $\pm$ 0.4   \\
\midrule[0pt]
\multicolumn{12}{c}{\textsc{Structural Graph Embedding Methods}} \\
\midrule[0pt]
\textbf{Struct2Vec~\cite{struc2vec}} & &    &
35.4 $\pm$ 1.0 & 37.6 $\pm$ 1.3 & 39.0 $\pm$ 1.1 &
31.2 $\pm$ 0.8 & 35.1 $\pm$ 0.9 & 36.5 $\pm$ 0.7 &
48.5 $\pm$ 0.3 & 49.2 $\pm$ 0.4 & 49.6 $\pm$ 0.4 \\
\textbf{GraphWave~\cite{graphwave}} &   &    &
39.5 $\pm$ 2.1 & 41.1 $\pm$ 1.5 & 42.2 $\pm$ 1.9 &
38.5 $\pm$ 1.2 & 40.6 $\pm$ 0.9 & 43.9 $\pm$ 1.0 & 
43.0 $\pm$ 2.0 & 43.3 $\pm$ 1.3 & 44.3 $\pm$ 1.5 \\
\textbf{DRNE~\cite{drne}} &   &    &
34.9 $\pm$ 1.5 & 36.5 $\pm$ 1.5 & 37.3 $\pm$ 1.6 &
30.8 $\pm$ 1.2 & 32.2 $\pm$ 1.2 & 34.6 $\pm$ 1.4 &
40.4 $\pm$ 0.7 & 41.6 $\pm$ 0.4 & 43.3 $\pm$ 0.5 \\
\midrule[0pt]
\multicolumn{12}{c}{\textsc{Standard Graph Neural Networks}} \\
\midrule[0pt]
\textbf{GCN~\cite{gcn}}  & \checkmark & \checkmark 
 & 81.6 $\pm$ 0.5 &	82.0 $\pm$ 0.4&	83.0 $\pm$ 0.5 &	
75.8 $\pm$ 0.5 & 	76.6 $\pm$ 0.3 &	76.8 $\pm$ 0.4 &
85.7 $\pm$ 0.7 &	86.1 $\pm$ 0.5&	86.4 $\pm$ 0.5 \\
\textbf{GAT~\cite{gat}}  & \checkmark   & \checkmark   
& 80.9 $\pm$ 0.7 &	81.4 $\pm$ 0.2	 &81.8 $\pm$ 0.5 &	
74.5 $\pm$ 0.7 &	75.5 $\pm$ 0.7 &	76.4 $\pm$ 0.5 &	
 83.3 $\pm$ 0.3 &	84.2 $\pm$ 0.3 &	84.3 $\pm$ 0.3   \\
\textbf{GraphSAGE~\cite{graphsage}} & \checkmark   & \checkmark   
& 81.3 $\pm$ 0.3 &	83.5 $\pm$ 0.3 &	84.2 $\pm$ 0.3 &	
72.9 $\pm$ 0.3 &	73.8 $\pm$ 0.2 &	76.4 $\pm$ 0.4 &	
86.6 $\pm$ 0.2 &	87.2 $\pm$ 0.3 &	88.0 $\pm$ 0.2 \\
\textbf{JKNet~\cite{jknet}} &\checkmark  & \checkmark  
 &81.3 $\pm$ 0.8&83.6 $\pm$ 0.8	&84.2 $\pm$ 0.8&
	71.5 $\pm$ 0.8&	72.5 $\pm$ 0.7	&73.3 $\pm$ 0.7&
		82.2 $\pm$ 0.4&	83.8 $\pm$ 0.5	&84.4 $\pm$ 0.4   \\
\textbf{DGI~\cite{dgi}}  & \checkmark  & 
&  76.2 $\pm$ 0.8& 	77.3 $\pm$ 0.9& 	78.2 $\pm$ 0.8& 	
74.5 $\pm$ 0.7& 	74.7 $\pm$ 0.7& 	75.4 $\pm$ 0.7& 	
78.2 $\pm$ 0.9& 	78.5 $\pm$ 0.9& 	79.5 $\pm$ 0.9 \\	
\midrule[0pt]
\multicolumn{12}{c}{\textsc{Structural Graph Neural Networks}} \\
\midrule[0pt]
\textbf{DemoNet~\cite{demonet}} & \checkmark   & \checkmark   
& 81.0 $\pm$ 0.6 &	82.4 $\pm$ 0.5 &	83.4 $\pm$ 0.7 &	
67.9 $\pm$ 0.7 &	68.5 $\pm$ 0.6 &	68.9 $\pm$ 0.6	 &
79.5 $\pm$ 0.4	 &80.5 $\pm$ 0.4	 &81.3 $\pm$ 0.4    \\
\textbf{Motif-CNN~\cite{motifcnn}} & \checkmark  & \checkmark  
 &81.6 $\pm$ 0.5 &	82.8 $\pm$ 0.5 &	83.2 $\pm$ 0.5 &	
73.4 $\pm$ 0.3 &	76.8 $\pm$ 0.3 &	77.1 $\pm$ 0.3 &	
87.3 $\pm$ 0.1 &	87.5 $\pm$ 0.1 &	88.2 $\pm$ 0.1    \\
\textbf{MCN~\cite{motif_attention_cikm19}} & \checkmark  & \checkmark  
 &81.1 $\pm$ 0.9 &	82.4 $\pm$ 0.8 &	83.1 $\pm$ 0.9 &	
73.2 $\pm$ 0.4 &	75.9 $\pm$ 0.7 &	76.6 $\pm$ 0.6 &	
85.2 $\pm$ 0.6 &	85.9 $\pm$ 0.5 &	86.4 $\pm$ 0.7    \\
\midrule[0pt]
\multicolumn{12}{c}{\textsc{Motif-regularized Graph Neural Networks~(\textbf{\name}) }} \\
\midrule[0pt]
\textbf{{InfoMotif-GCN}}  & \checkmark & \checkmark & 
 \textbf{85.7 $\pm$ 0.4} &	\textbf{87.4 $\pm$ 0.4}&	\textbf{88.2 $\pm$ 0.3}&	
\textbf{77.7 $\pm$ 0.5}&	\textbf{78.5 $\pm$ 0.5}&	\textbf{80.1 $\pm$ 0.5}&	
\textbf{87.5 $\pm$ 0.2}&	\textbf{88.3 $\pm$ 0.2}&	\textbf{88.7 $\pm$ 0.4}   \\
\textbf{{InfoMotif-JKNet}}  & \checkmark & \checkmark & 
85.5 $\pm$ 0.3 &	86.5 $\pm$ 0.5&	88.0 $\pm$ 0.2&	
74.5 $\pm$ 0.8&	 76.7 $\pm$ 0.9&	77.8 $\pm$ 0.9&	
87.0 $\pm$ 0.2&	 87.9 $\pm$ 0.3&	88.2 $\pm$ 0.3   \\
\textbf{{InfoMotif-GAT}} & \checkmark & \checkmark & 
85.5 $\pm$ 0.3 & 87.2 $\pm$ 0.7 & 88.0 $\pm$ 0.2 &
76.5 $\pm$ 0.5 & 77.0 $\pm$ 0.4 & 78.9 $\pm$ 0.4 &
85.9 $\pm$ 0.4 & 86.2 $\pm$ 0.5 & 86.3 $\pm$ 0.5\\

\bottomrule
\end{tabular}
\caption{Node classification results (\% test accuracy) on citation networks using 10 random train/validation/test splits per training ratio (20\%, 40\% and 60\%). 
$\mathbf{X}$ and $\mathbf{Y}$ denote the use of node attributes and training labels respectively towards representation learning.
We report mean accuracy and standard deviation over 5 trials. We show GraphSAGE results with the best performing aggregator.~\name~consistently improves results of all three base GNNs by 3.5\% on average across datasets.}
\vspace{-10pt}
\label{tab:citation_results}
\end{table*}

\subsection{\textbf{Datasets}}
\label{sec:datasets}
We conduct experiments on two diverse types of benchmark datasets: \textit{citation networks} that exhibit strong homophily and \textit{air-traffic networks} that depend on structural roles (Table~\ref{tab:stats}).
\begin{itemize}[leftmargin=*]
\item \textbf{Citation Networks:} We consider three benchmark datasets, Cora, Citeseer, and PubMed~\cite{citation}, where nodes correspond to documents and edges represent citation links. Each document is associated with a bag-of-words feature vector and the task is to classify documents into different research topics.
\item \textbf{Air-Traffic Networks:} We use three undirected networks Brazil, Europe, and USA~\cite{struc2vec} where nodes correspond to airports and edges indicate the existence of commercial flights.
Class labels are assigned based on activity level, measured by the cardinality of flights or people that passed the airports. We use one-hot indicator vectors as node attributes.
Notice that class labels are related to the role played by airports.
\end{itemize}

\subsection{\textbf{Baselines}}
\label{sec:baselines}
We organize competing baselines into four categories based on whether they are \textit{proximity-based vs. structural}; and the paradigm of \textit{embedding learning vs. graph neural networks}:

\begin{itemize}[leftmargin=*]

    \item \textbf{Proximity-based embedding methods}: Conventional methods, node2vec~\cite{node2vec} that learns from second-order random walks, and motif2vec~\cite{motif2vec} that models higher-order proximity.
    
    \item \textbf{Structural embedding methods}: Structural role-aware models struc2vec~\cite{struc2vec}, GraphWAVE~\cite{graphwave}, and DRNE~\cite{drne}.
    
    \item \textbf{Standard Graph Neural Networks}: State-of-the-art GNN models based on localized message passing: GCN~\cite{gcn}, GraphSAGE~\cite{graphsage}, GAT~\cite{gat}, JK-Net~\cite{jknet} and DGI~\cite{dgi}.
    
    \item \textbf{Structural Graph Neural Networks}: Motif-based Motif-CNN~\cite{motifcnn}, MCN~\cite{motif_attention_cikm19} and degree-specific DEMO-Net~\cite{demonet}.
    
\end{itemize}

\subsection{\textbf{Experimental Setup}}
We tested~\name~by integrating GCN, JK-Net and GAT as base GNNs within our framework. %
We consider the largest connected component in each dataset, and use the set of all directed 3-node motifs in citation networks and undirected 3-node motifs in air-traffic networks (\Cref{fig:motifs}).
To fairly compare different models~\cite{pitfalls}, we evaluate different train/validation/test splits (training ratios of 20\%, 40\%, and 60\%).
We create 10 random data splits per training ratio and report the mean test classification accuracy along with standard deviation.

All experiments were conducted on a Tesla K-80 GPU using PyTorch. Our implementation of~\name~is publicly available\footnote{
https://github.com/CrowdDynamicsLab/InfoMotif}.
For citation networks, we use two-layer base GNNs with layer sizes of 256 each, while using 64 for the smaller air-traffic networks.
We train the base JK-Net using 4 GCN layers and maxpool layer aggregation, while the base GAT learns 8 attention heads per layer.
The model is trained for a maximum of 100 epochs with a batch size of 256 nodes with Adam optimizer.
We also apply dropout with a rate of 0.5, and tune the learning rate in the range $\{ 10^{-4}, 10^{-3}, 10^{-2}\}$.

\label{sec:setup}

\begin{table*}[t]
\centering
\noindent\setlength\tabcolsep{2.9pt}
\begin{tabular}{@{}p{0.13\linewidth}@{\hspace{0pt}}
K{\factortraining\linewidth}K{\factortraining\linewidth}@{\hspace{0pt}}
K{\factor\linewidth}K{\factor\linewidth}K{\factor\linewidth}@{\hspace{10pt}}
K{\factor\linewidth}K{\factor\linewidth}K{\factor\linewidth}@{\hspace{10pt}}
K{\factor\linewidth}K{\factor\linewidth}K{\factor\linewidth}@{}} \\
\toprule
\multirow{1}{*}{\textbf{}} &  \multicolumn{2}{c}{\textbf{Data}} & \multicolumn{3}{c}{\textbf{USA}}  & \multicolumn{3}{c}{\textbf{Europe}} & \multicolumn{3}{c}{\textbf{Brazil}} \\
\cmidrule{2-3} \cmidrule(lr){4-6} \cmidrule(lr){7-9} \cmidrule(lr){10-12}
\multirow{1}{*}{\textbf{Training Ratio}} & \textbf{X} & \textbf{Y} & \textbf{20\%} & \textbf{40\%} & \textbf{60\%} & \textbf{20\%} & \textbf{40\%} & \textbf{60\%} & \textbf{20\%} & \textbf{40\%}  & \textbf{60\%} \\
\midrule
\multicolumn{12}{c}{\textsc{Proximity-based Graph Embedding Methods}} \\
\midrule[0pt]
\textbf{{Node2Vec}~\cite{node2vec}} &   &
 & 24.6 $\pm$ 0.9	&24.8 $\pm$ 0.9	&25.6 $\pm$ 0.9	&
36.5 $\pm$ 1.0&	37.4 $\pm$ 1.1&	38.0 $\pm$ 1.0	&
26.3 $\pm$ 1.4	&30.4 $\pm$ 1.3	&33.9 $\pm$ 1.4    \\
\textbf{{Motif2Vec}~\cite{motif2vec}}  &   &
 & 51.3 $\pm$ 1.1 	&54.8 $\pm$ 1.1	&55.0 $\pm$ 1.1	&
37.1 $\pm$ 1.2&	38.1 $\pm$ 1.2&	39.5 $\pm$ 1.1	&
27.2 $\pm$ 1.5	&33.9 $\pm$ 1.5	&35.7 $\pm$ 1.5  \\
\midrule[0pt]
\multicolumn{12}{c}{\textsc{Structural Graph Embedding Methods}} \\
\midrule[0pt]
\textbf{{Struct2Vec}~\cite{struc2vec}}  &  &
 & 50.4 $\pm$ 0.8&51.3 $\pm$ 0.8	&53.8 $\pm$ 0.8	&
42.5 $\pm$ 0.7&45.6 $\pm$ 0.8&48.8 $\pm$ 0.7	&
45.8 $\pm$ 1.1&51.8 $\pm$ 1.1	&57.1 $\pm$ 1.1   \\
\textbf{GraphWave~\cite{graphwave}} &   &    &
45.2 $\pm$ 1.4 & 48.0 $\pm$ 1.4 & 51.4 $\pm$ 1.5 & 
38.1 $\pm$ 1.9 & 41.1 $\pm$ 1.6 & 42.1 $\pm$ 2.0 & 
40.2 $\pm$ 2.0 & 43.1 $\pm$ 1.8 & 48.5 $\pm$ 2.2 \\

\textbf{DRNE~\cite{drne}} &  &    &
51.3 $\pm$ 1.1 & 52.4 $\pm$ 1.1 & 53.3 $\pm$ 1.1 &
43.1 $\pm$ 1.7 & 47.6 $\pm$ 1.3 & 50.8 $\pm$ 1.6 & 
46.5 $\pm$ 2.7 & 50.2 $\pm$ 2.3 & 58.1 $\pm$ 2.0 \\

\midrule[0pt]
\multicolumn{12}{c}{\textsc{Standard Graph Neural Networks}} \\
\midrule[0pt]
\textbf{GCN~\cite{gcn}}  & \checkmark  & \checkmark  
& 51.9 $\pm$ 0.9	&56.0 $\pm$ 0.9&	57.0 $\pm$ 0.8&
37.4 $\pm$ 0.9&40.1 $\pm$ 0.8	&41.0 $\pm$ 0.8&
	36.5 $\pm$ 1.5	&38.9 $\pm$ 1.6	&39.3 $\pm$ 1.4  \\
\textbf{GAT~\cite{gat}}  & \checkmark   & \checkmark   
& 52.7 $\pm$ 1.0	&53.5 $\pm$ 0.9 &56.3 $\pm$ 0.9	&
31.5 $\pm$ 1.0	&34.3 $\pm$ 1.0	&38.0 $\pm$ 1.0	&
37.3 $\pm$ 1.6	&37.9 $\pm$ 1.6	&38.2 $\pm$ 1.7  \\
\textbf{GraphSAGE~\cite{graphsage}} & \checkmark  & \checkmark   
& 45.3 $\pm$ 1.2	&49.4 $\pm$ 1.2&	50.4 $\pm$ 1.1	&
28.8 $\pm$ 1.0&	32.5 $\pm$ 1.0	&37.9 $\pm$ 1.0&
	36.1 $\pm$ 1.6	&37.5 $\pm$ 1.6	&39.3 $\pm$ 1.7 \\

\textbf{JKNet~\cite{jknet}} &\checkmark  & \checkmark  
 &53.8 $\pm$ 1.2&56.1 $\pm$ 1.0	&61.3 $\pm$ 1.0&
	49.7 $\pm$ 1.1&	53.8 $\pm$ 1.1	&54.3 $\pm$ 1.2&
		55.9 $\pm$ 1.5&	58.4 $\pm$ 1.8	&60.0 $\pm$ 1.4  \\
\textbf{DGI~\cite{dgi}}  & \checkmark   &
&  46.4 $\pm$ 1.3&	47.3 $\pm$ 1.2&48.1 $\pm$ 1.2	&
37.5 $\pm$ 1.5	&39.9 $\pm$ 1.5&	42.3 $\pm$ 1.4	&
41.4 $\pm$ 1.6	&45.2 $\pm$ 1.7	&44.1 $\pm$ 1.5\\

\midrule[0pt]
\multicolumn{12}{c}{\textsc{Structural Graph Neural Networks}} \\
\midrule[0pt]

\textbf{DemoNet~\cite{demonet}} & \checkmark   & \checkmark  
 & 58.6 $\pm$ 1.2	&58.8 $\pm$ 1.1&	61.3 $\pm$ 1.0	&
40.4 $\pm$ 1.3&46.2 $\pm$ 1.2	&47.5 $\pm$ 1.2	&
46.1 $\pm$ 1.4	&48.9 $\pm$ 1.5	&49.2 $\pm$ 1.5  \\
\textbf{Motif-CNN~\cite{motifcnn}} & \checkmark  & \checkmark  
 &53.6 $\pm$ 1.0&54.2 $\pm$ 1.0	&55.6 $\pm$ 0.9&
	37.9 $\pm$ 1.0&	41.1 $\pm$ 1.1	&42.8 $\pm$ 1.0&
	28.9 $\pm$ 1.6&	35.7 $\pm$ 1.7	&39.3 $\pm$ 1.7   \\
\textbf{MCN~\cite{motif_attention_cikm19}} & \checkmark  & \checkmark  
 &54.8 $\pm$ 1.4&54.9 $\pm$ 1.3	&55.3 $\pm$ 1.1&
	36.8 $\pm$ 1.2&	39.6 $\pm$ 1.5	&41.2 $\pm$ 1.4&
	42.9 $\pm$ 1.6&	43.6 $\pm$ 1.4	&47.2 $\pm$ 1.5   \\

\midrule[0pt]
\multicolumn{12}{c}{\textsc{Motif-regularized Graph Neural Networks~(\textbf{\name}) }} \\
\midrule[0pt]

\textbf{{InfoMotif-GCN}}  & \checkmark  & \checkmark &
59.5 $\pm$ 0.9	&62.9 $\pm$ 0.7	&65.0 $\pm$ 0.7&
\textbf{53.5 $\pm$ 0.6}	&56.9 $\pm$ 0.6&\textbf{58.8 $\pm$ 0.7}&
	56.6 $\pm$ 1.2	&60.7 $\pm$ 1.2	&67.9 $\pm$ 1.1  \\
\textbf{{InfoMotif-JKNet}}  & \checkmark & \checkmark & 
 \textbf{61.8 $\pm$ 1.6} &	\textbf{64.3 $\pm$ 1.2}&	\textbf{67.5 $\pm$ 1.5}&	
53.1 $\pm$ 1.2&	\textbf{56.9 $\pm$ 0.6}&	57.5 $\pm$ 1.2&	
\textbf{62.7 $\pm$ 1.8}&	\textbf{67.9 $\pm$ 1.5}&	\textbf{80.4 $\pm$ 1.9}   \\
\textbf{{InfoMotif-GAT}} & \checkmark & \checkmark & 
58.0 $\pm$ 0.4 & 60.4 $\pm$ 0.3 & 62.6 $\pm$ 0.7 &
 46.0 $\pm$ 1.5 & 50.0 $\pm$ 2.0 & 56.3 $\pm$ 0.5 &
50.6 $\pm$ 1.3 & 56.3 $\pm$ 1.1 & 58.9 $\pm$ 1.3\\
\bottomrule
\end{tabular}
\caption{Node classification results (\% test accuracy) on air-traffic networks. Structural embedding methods and GNNs typically outperform proximity-based models.~\name~JK-Net achieves significant gains of 4\% to 14\% across datasets.
}
\vspace{-10pt}
\label{tab:airport_results}
\end{table*}

\subsection{\textbf{Experimental Results}}
\label{sec:results}

Our experimental results comparing~\name~with three base GNNs, against competing baselines on citation and air-traffic networks, are shown in Tables~\ref{tab:citation_results} and~\ref{tab:airport_results} respectively.

In citation networks, GNNs generally outperform conventional methods.
Moreover, attribute-agnostic structural embedding methods perform poorly and structural GNNs perform comparably to standard GNNs.
Citation networks exhibit strong attribute homophily in local neighborhoods;
thus, structural GNNs do not provide much benefits over state-of-the-art message-passing GNNs.
In contrast, our framework~\name~regularizes GNNs to discover distant nodes with similar attributed structures across the entire graph.
~\name~achieves consistent average accuracy gains of 3\%  for all three variants. %

In air-traffic networks, structural embedding methods outperform their proximity-based counterparts, with a similar trend for structural GNNs.
Here, class labels rely more on node structural roles than the labels of neighbors.
JK-net outperforms competing GNNs, signifying the importance of long-range dependencies in air-traffic networks.
~\name~enables GNNs to learn structural roles agnostic to network proximity,
and achieves significant gains of 10\% on average across all datasets.

\subsection{\textbf{Ablation Study}}
\label{sec:ablation}
We present an ablation study on citation networks to analyze the importance of major components in~\name~(Table~\ref{tab:ablation})

\begin{itemize}[leftmargin=*]
    \item \textbf{Remove novelty-driven sample weighting}. 
    We set the novelty $\beta_v=1$ (\Cref{eqn:supervised_loss}) to test the importance of addressing motif occurrence skew. We observe consistent 1\% gains due to our novelty-driven sample weighting.
    \item \textbf{Remove task-driven motif weighting}. 
    We remove the node-sensitive motif weights from the motif regularization loss (\Cref{eqn:full_mi_loss}) by setting $\alpha_{vt}=1$ for every node-motif pair.
    Contextually weighting different motif regularizers at a node-level granularity results in 2\% average accuracy gains.
    \item \textbf{Remove both novelty and task driven weighting}.
This variant applies a uniform motif regularization over all nodes without  distinguishing the nodes-sensitive relevance of each motif, which significantly degrades classification accuracy. %
\end{itemize}

\renewcommand*{\factor}{0.142}
\begin{table}[hbtp]
    \centering
    \noindent\setlength\tabcolsep{2.9pt}
\begin{tabular}{@{}p{0.5\linewidth}@{\hspace{9pt}}K{\factor\linewidth}K{\factor\linewidth}K{\factor\linewidth}@{}}
        \toprule
        \textbf{Dataset} & \textbf{Cora} & \textbf{Citeseer} & \textbf{Pubmed} \\ 
        \midrule
        InfoMotif-GCN ($L_S + \lambda L_{MI}$) &  \textbf{87.4 $\pm$ 0.4} & \textbf{78.5 $\pm$ 0.5} & \textbf{88.3 $\pm$ 0.2}  \\
        w/o novelty weights ($\beta_v=1$ in~\cref{eqn:supervised_loss}) & 86.4 $\pm$ 0.5  & 77.6 $\pm$ 0.5 & 87.8 $\pm$ 0.3\\
        w/o task weights ($\alpha_{vt} = 1$ in~\cref{eqn:full_mi_loss}) & 84.6 $\pm$ 0.4 & 77.3 $\pm$ 0.4 & 87.3 $\pm$ 0.2\\ 
        w/o novelty and task weights & 84.0 $\pm$ 0.5 & 76.4 $\pm$ 0.6 & 87.3 $\pm$ 0.2\\
        Base model GCN  ($L_B$) & 82.0 $\pm$ 0.4 &  76.6 $\pm$ 0.3 & 86.1 $\pm$ 0.5 \\
        \bottomrule
    \end{tabular}
    \caption{Ablation study results with 40\% training ratio on citation networks.
    The novelty and task weighting strategies improve classification accuracies by 2\% on average.}
    \label{tab:ablation}
\end{table}

\subsection{\textbf{Qualitative Analysis}}
\label{sec:analysis}
We qualitatively examine the source of~\name's gains over the base GNN (GCN due to its consistent performance).
by analyzing \textit{label sparsity} and \textit{attribute diversity} in local node neighborhoods, on the Cora and Citeseer citation networks.

\subsubsection*{\textbf{Label Sparsity}}
\begin{figure}[t]
    \centering
    \includegraphics[width=\linewidth]{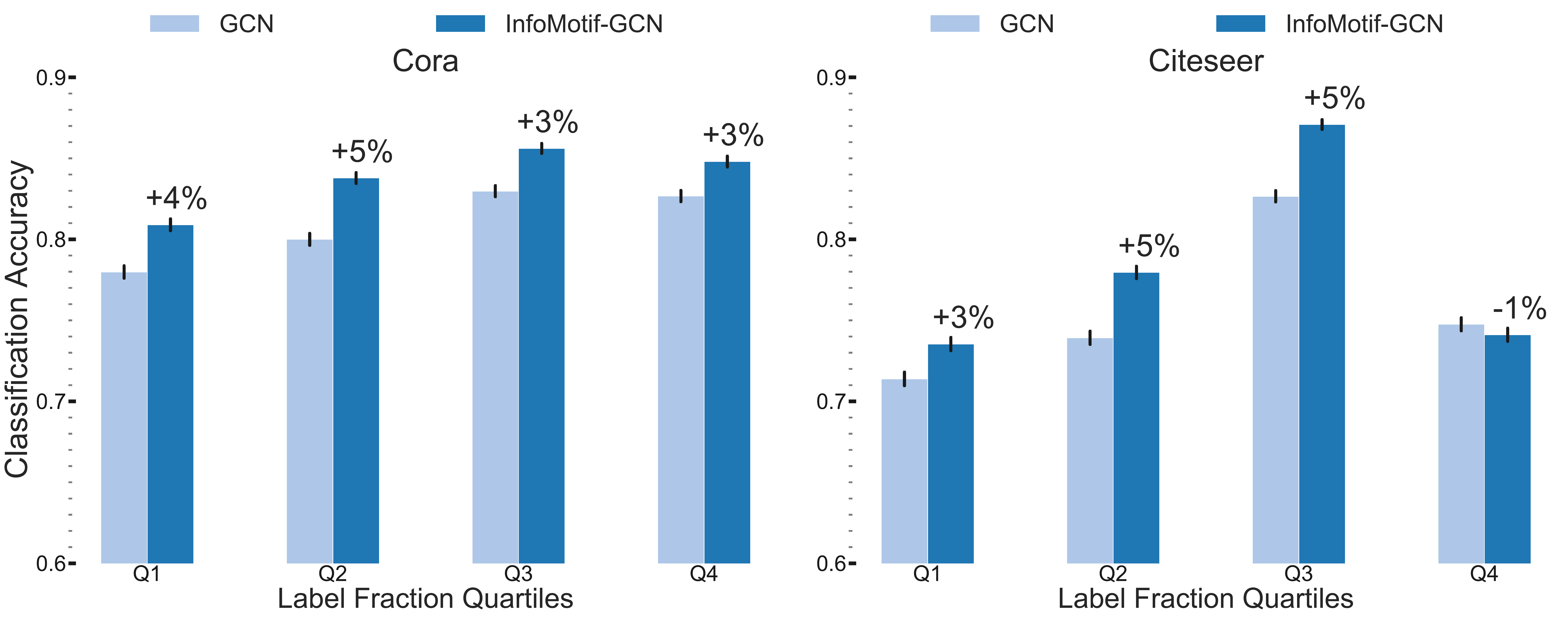}
    \caption{Classification accuracy over \textit{label fraction} quartiles. (Q1: smaller label fraction).~\name~has larger gains over GCN in Q1 \& Q2 (nodes that exhibit label sparsity)}
    \label{fig:label_sparsity}
\end{figure}

We define the \textit{label fraction} for a node as the fraction of labeled training nodes in its 2-hop neighborhood, \textit{i.e.}, a node exhibits label sparsity if it has very few or no labeled training nodes within its 2-hop aggregation range.
We separate test nodes into four quartiles by their label fraction.
Figure~\ref{fig:label_sparsity} depicts classification results for GCN and~\name-GCN under each quartile (Q1 has nodes with small label fractions).

~\name~has stronger performance gains over GCN for nodes with smaller label fractions (quartiles Q1 and Q2), which empirically validates the efficacy of our motif-based regularization framework in addressing the key limitation of GNNs~(\Cref{sec:role_learning}), \textit{i.e.},~\name~benefits nodes with very few or no labeled nodes within their $k$-hop aggregation ranges.

\subsubsection*{\textbf{Attribute Diversity}}
\begin{figure}[t]
    \centering
    \includegraphics[width=\linewidth]{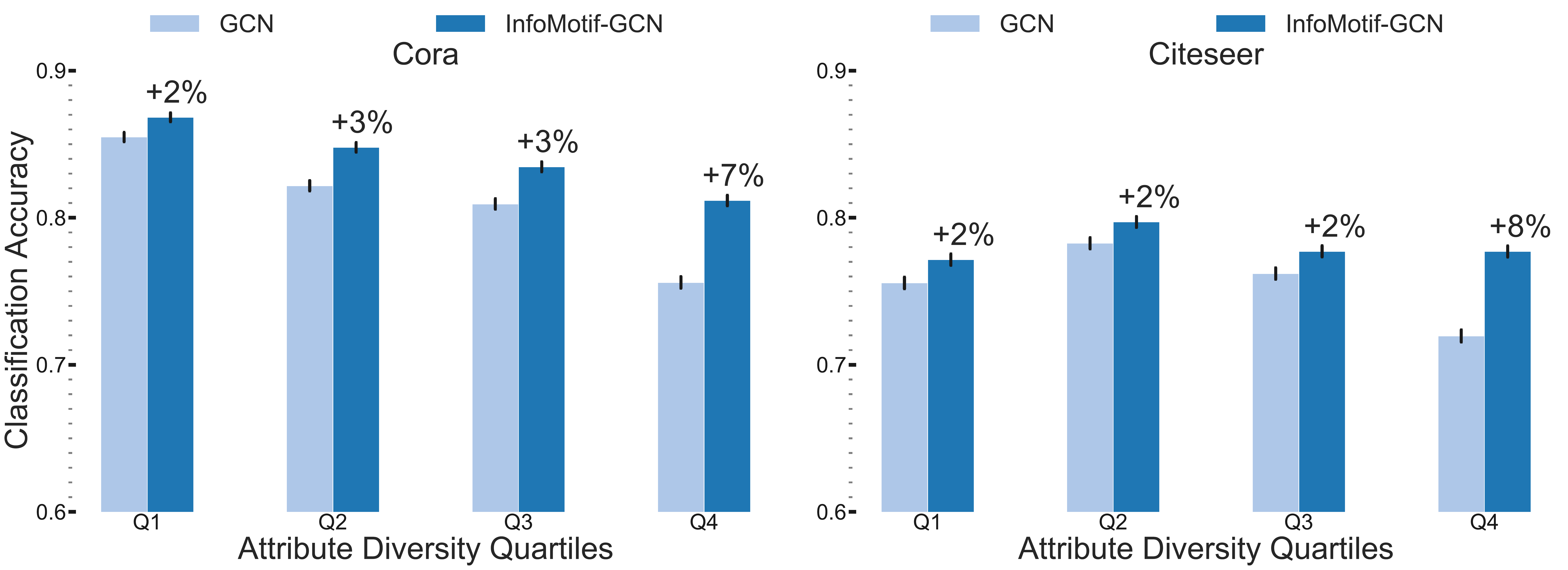}
    \caption{Classification accuracy across attribute diversity quartiles. (Q4: high attribute diversity).~\name~has stronger gains in Q3 \& Q4 (nodes with diverse attributed neighborhoods).}
    \label{fig:cosine_sim}
\end{figure}

We measure the local \textit{attribute diversity} of a node by the mean pair-wise attribute dissimilarity (computed by cosine distance) of itself with other nodes in its 2-hop neighborhood, \textit{i.e.}, a node that exhibits strong homophily with its neighbors has low attribute diversity. 
We report classification results across attribute diversity quartiles in Figure~\ref{fig:cosine_sim}.

Nodes with diverse attributed neighborhoods are typically harder examples for classification.
Regularizing GNNs to learn attributed structures via motif occurrences can accurately classify diverse nodes, as evidenced by the higher relative gains of~\name~for diverse nodes (quartiles Q3 and Q4).

\subsection{\textbf{Parameter Sensitivity}}
\label{sec:sensitivity}
\begin{figure}[hbtp]
    \centering
    \includegraphics[width=\linewidth]{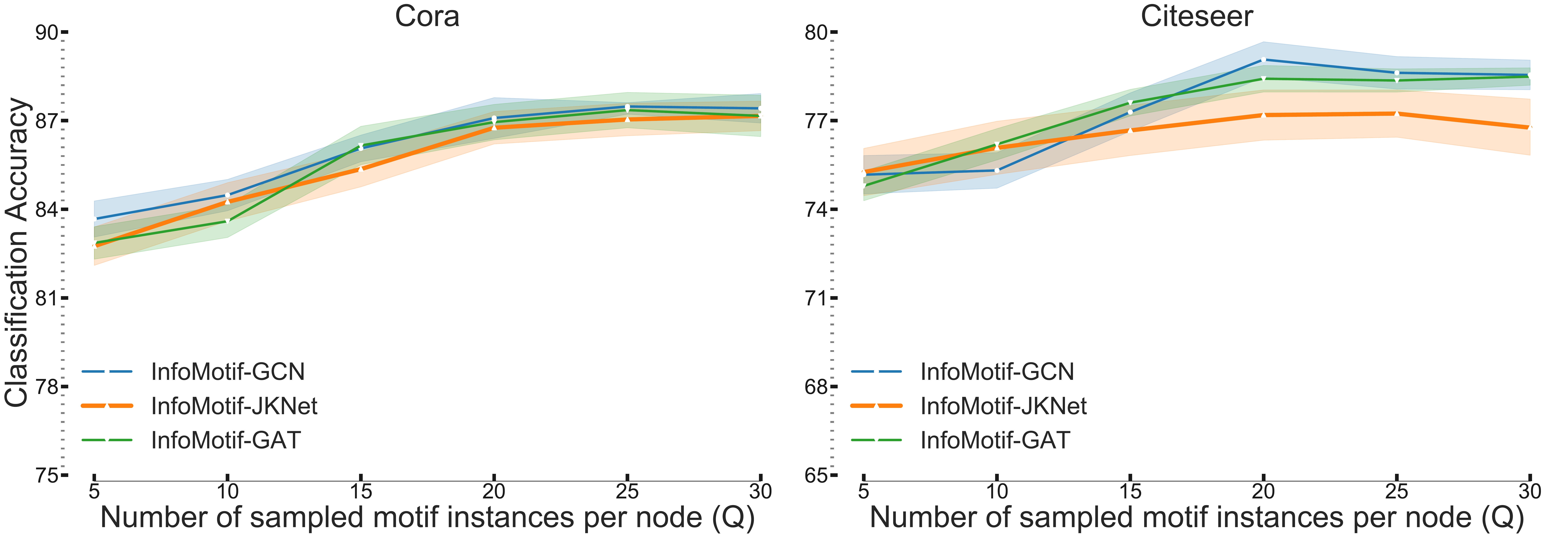}
    \caption{Classification accuracy increases slowly with the number of sampled motif instances and stabilizes around 15 to 20. Variance bands indicate 95\% confidence intervals over 10 runs.}
    \label{fig:sensitivity}
\end{figure}

We examine the effect of hyper-parameter $Q$ that controls the number of motif instances sampled per node to train our motif-based discriminators (\Cref{eqn:mi_loss}).~\Cref{fig:sensitivity} shows variation in accuracies of our three GNN variants with the number of sampled instances (5 to 30), on Cora and Citeseer networks.

Performance of all GNN variants stabilize with 20 instances across both datasets.
Since the complexity of our framework scales linearly with $Q$, we fix $Q=20$ across datasets to provide an effective trade-off between compute-cost and performance

\subsection{\textbf{Efficiency Analysis}}
\label{sec:efficiency}
\begin{figure}[hbtp]
    \centering
    \includegraphics[width=0.9\linewidth]{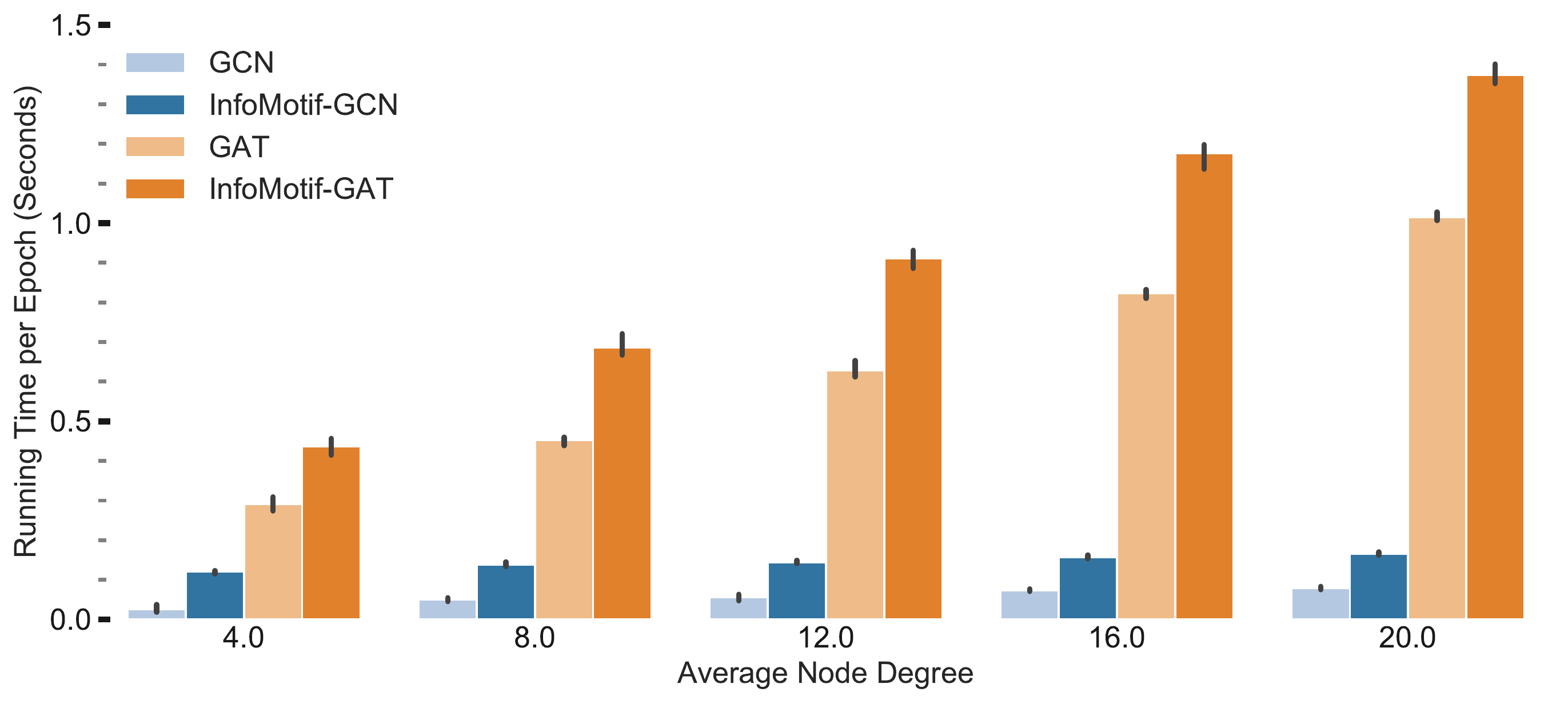}
    \caption{Runtime comparison of~\name~variants with its base GNNs.~\name~has minimal computational overheads; notice the nearly constant runtime gap with increasing node degree.}
    \label{fig:runtime}
\end{figure}

We empirically evaluate the added complexity of~\name~on two GNN models, GCN and GAT.
We report the time per epoch %
on synthetically generated Barabasi-Albert networks~\cite{barabasi} with 5000 nodes and increasing link density (\Cref{fig:runtime}).

\name~adds a small fraction of the base GNN runtime, and the added complexity scales linearly with the number of nodes, as evidenced by its nearly constant runtime gap over increasing link density (\Cref{fig:runtime}).
Furthermore, our GCN variant~\name-GCN is significantly more efficient than GAT.

\subsection{\textbf{Discussion}}
\label{sec:discussion}
Our framework is orthogonal to advances in GNN architectures that enhance the structural distinguishability of node representations through carefully designed aggregators.
We regularize arbitrary GNNs to learn statistical correspondences between distant nodes with co-varying attribute structures.
Our abstraction of roles through motifs generalizes across diverse types of networks, \textit{e.g.}, signed and heterogeneous motifs~\cite{heterogeneous_motif}.

Our key hypothesis is the importance of attribute co-variance in local structures towards the learning application (\textit{e.g.}, classification in social networks).
Our substantial gains on two diverse classes of datasets indicates broad applicability for~\name~across networks with varied structural characteristics.
However, the gains may diminish in application scenarios where learning such co-variance is not beneficial or even necessary.

\section{Related Work}
GNNs learn node representations by recursively aggregating features from local neighborhoods in an end-to-end manner, with diverse applications, including information diffusion prediction~\cite{infvae}, social recommendation~\cite{social_adv}, and community question answering~\cite{irgcn}.
Graph Convolutional Networks (GCNs)~\cite{gcn} learn degree-weighted aggregators, which can be interpreted as a special form of Laplacian smoothing~\cite{gcn_oversmoothing}.
Many models generalize GCN with a wide range of aggregators, \textit{e.g.}, self-attentions~\cite{gat, dysat}, mean and max pooling functions~\cite{graphsage}, etc.
However, all these models learn node representations that inherently overfit to the $k$-hop neighborhood around each node.

There are two broad categories of techniques that capture contributions from distant nodes for graph representation learning: \textit{non-local GNNs}, and \textit{structural role-based embeddings}.

Non-local methods expand the propagation range of GNNs to aggregate node representations of differing localities, \textit{e.g.}, JKNet~\cite{jknet} uses skip-connections to vary the influence radius per node, PGNN~\cite{pgnn} captures global network positions via shortest-paths, and DGI~\cite{dgi} maximizes MI between node representations and a summary representation of the entire graph.
However, they either operate on a local scale~\cite{gmi}, or learn coarse structural properties,  which limits their ability to capture features from distant yet structurally similar nodes.

Role-aware models embed structurally similar nodes close in the latent space, independent of network position~\cite{rolx, rossi2019community}.
A few approaches~\cite{drne} employ strict definitions of structural equivalence to embed nodes with identical local structures to the same point in the latent space, while others utilize structural node features (\textit{e.g.}, node degrees, motif count statistics) to extend classical proximity-preserving embedding methods, \textit{e.g.}, feature-based matrix factorization~\cite{hone} and random walk methods~\cite{struc2vec}.
Notably, a few methods design structural GCNs via motif adjacency matrices~\cite{motif_attention_cikm19, motifcnn, metagnn}.
However, all these methods model structural roles without considering node attributes.
~\name~is different since we regularize GNNs based on the co-variance of attributes and motif structures.

A related direction is higher-order network representation learning that models proximity via network motifs~\cite{motif2vec}. %
However, such representations are still highly localized and cannot identify structurally similar nodes independent of network proximity.
In contrast, we %
contrastively learn attribute correlations in motifs to identify correspondences between distant nodes. %

\section{Conclusion}
\label{sec:conclusion}
This paper presents a new class of motif-regularized GNNs with an architecture-agnostic framework~\name~for semi-supervised learning on graphs.
To overcome limitations of prior GNNs due to localized message passing,
we introduce attributed structural roles to regularize GNNs by learning statistical dependencies between structurally similar nodes with co-varying attributes, independent of network proximity.
~\name~maximizes motif-based mutual information, and dynamically prioritizes the significance of different motifs.
Our experiments on six real-world datasets show substantial consistent gains for~\name~over state-of-the-art methods.

\bibliographystyle{IEEEtran}
\bibliography{main}

\end{document}